\documentclass[epjCONF, onecolumn]{svjour}
\usepackage{graphics}
\usepackage[varg]{txfonts} % Times fonts
\usepackage[latin1]{inputenc}
\session-title{Conference Title, to be filled}
\begin{document}
\title{Stellar dynamics around transient co-rotating spiral arms}
\author{Daisuke Kawata\thanks{\email{d.kawata@ucl.ac.uk}} \and Robert J.J. Grand
 \and Mark Cropper}
\institute{Mullard Space Science Laboratoy, University College London, Holmbury St. Mary, Dorking, Surrey, RH5 6NT, UK}
\abstract{
Spiral density wave theory attempts to describe the spiral pattern in spiral galaxies
in terms of a long-lived wave structure with a constant pattern speed in order to
avoid the winding dilemma. The pattern is consequently a rigidly rotating, long-lived
feature. We run an N-body/SPH simulation of a Milky Way-sized barred disk, and find that the spiral arms are transient
features whose pattern speeds decrease with radius, in such a way that the pattern
speed is almost equal to the rotation curve of the galaxy. We trace particle motion around the spiral arms. 
We show that particles from behind and in front of the spiral arm are drawn towards and join the arm. Particles
move along the arm in the radial direction and we find a clear trend that they migrate toward the outer (inner) radii
on the trailing (leading) side of the arm. Our simulations demonstrate that at all radii where there is a co-rotating spiral arm the particles continue to be accelerated (decelerated) by the spiral arm for long periods, which leads to strong migration.
} %end of abstract
\maketitle
\section{Introduction}
\label{intro-sec}

Spiral density wave theory suggested by \cite{ls64} has been the most widely accepted theory of the spiral arm for almost
50 years. This theory considers the spiral arm to be the crest of a stellar density wave that rotates around the centre of
the Galaxy at an angular pattern speed, ${\rm \Omega}_{\rm p}$, that does not vary with galacto-centric radius.
Spiral density wave theory naturally explains the long-lived nature of the spiral arm.
However, there are no numerical simulations that reproduce long-lived spiral arms
like those predicted by spiral density wave theory (e.g. \cite{js11}).
Recently, there have been observational studies with evidence of a pattern speed that varies with radius,
which is different from the constant pattern speed of spiral density wave theory \cite{mrm06,frdlw11}.
We analyse the pattern speed of the spiral arms and the dynamics of particles around the arm using
an N-body/SPH simulation of a Milky Way-sized barred disk.

\section{Model and Results}
\label{mod-sec}

 We use our original N-body/SPH chemodynamical simulation code, GCD+
\cite{kg03}. The new version of GCD+ includes modeling of
powerful supernovae feedback and metal diffusion of the gas. We have set
up initially an equilibrium gas and stellar disks embedded in a fixed DM halo
potential, and follow 2 Gyr evolution. The mass and initial scale length of the
gas and stellar disks are $10^{10}$ M$_{\odot}$ and $4\times10^{10}$ M$_{\odot}$, and 4 and 2.5 kpc
respectively. The particle mass for both gas and star is $10^5$ M$_\odot$.

\begin{figure}
\resizebox{0.6\columnwidth}{!}{ \includegraphics{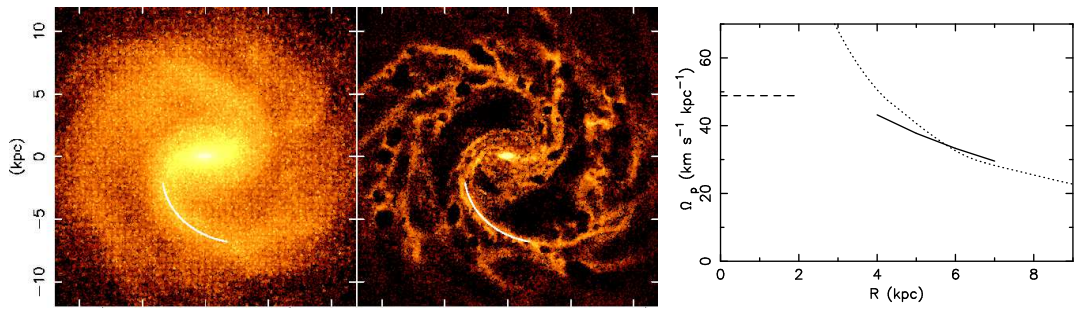} }
\resizebox{0.4\columnwidth}{!}{ \includegraphics{kawata_d_fig1b.eps} }
\caption{Face-on view snapshot of stellar ({\it Left}) and gas ({\it Middle}) disk of the
simulated galaxy. The white lines show our fitting of a spiral arm.
{\it Right}: The pattern speed of spiral arm (solid line), and stellar rotation speed (dotted line).
The pattern speed of the central bar is also shown with dashed line.
%{\it Right}: The change in angular momentum over 80 Myr as a function of initial angular momentum,
%L$_{\rm ini}$, of all star particles. Overplotted in black dots are the selected particle in Fig. 2.
%Red (positive migrators) and blue (negative migrators) dots are the particles who experienced highest gain and lose of
%angular momentum.
}
\label{fig:1}       % Give a unique label
\end{figure}

\begin{figure}
\resizebox{0.33\columnwidth}{!}{ \includegraphics{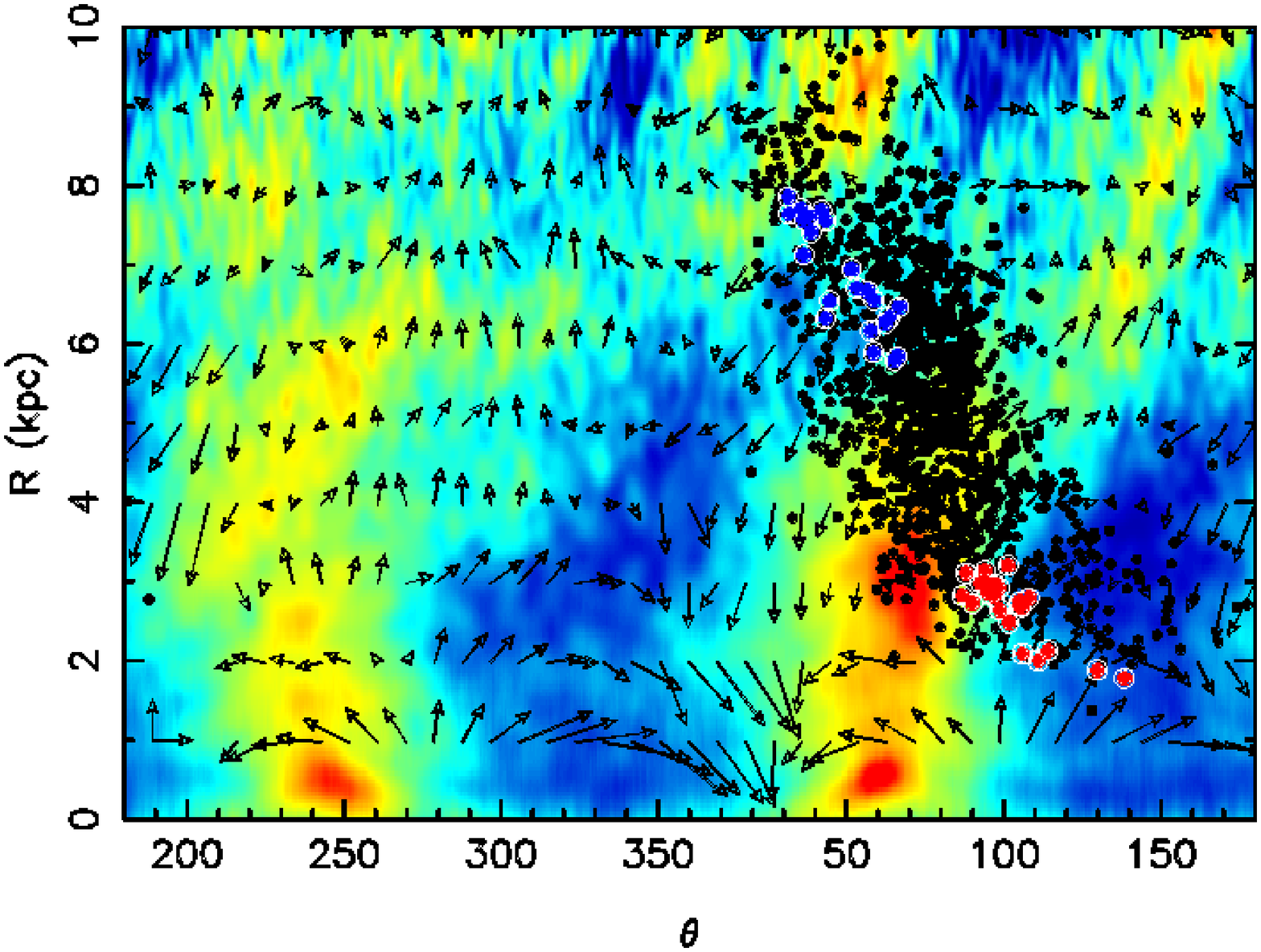} }
\resizebox{0.33\columnwidth}{!}{ \includegraphics{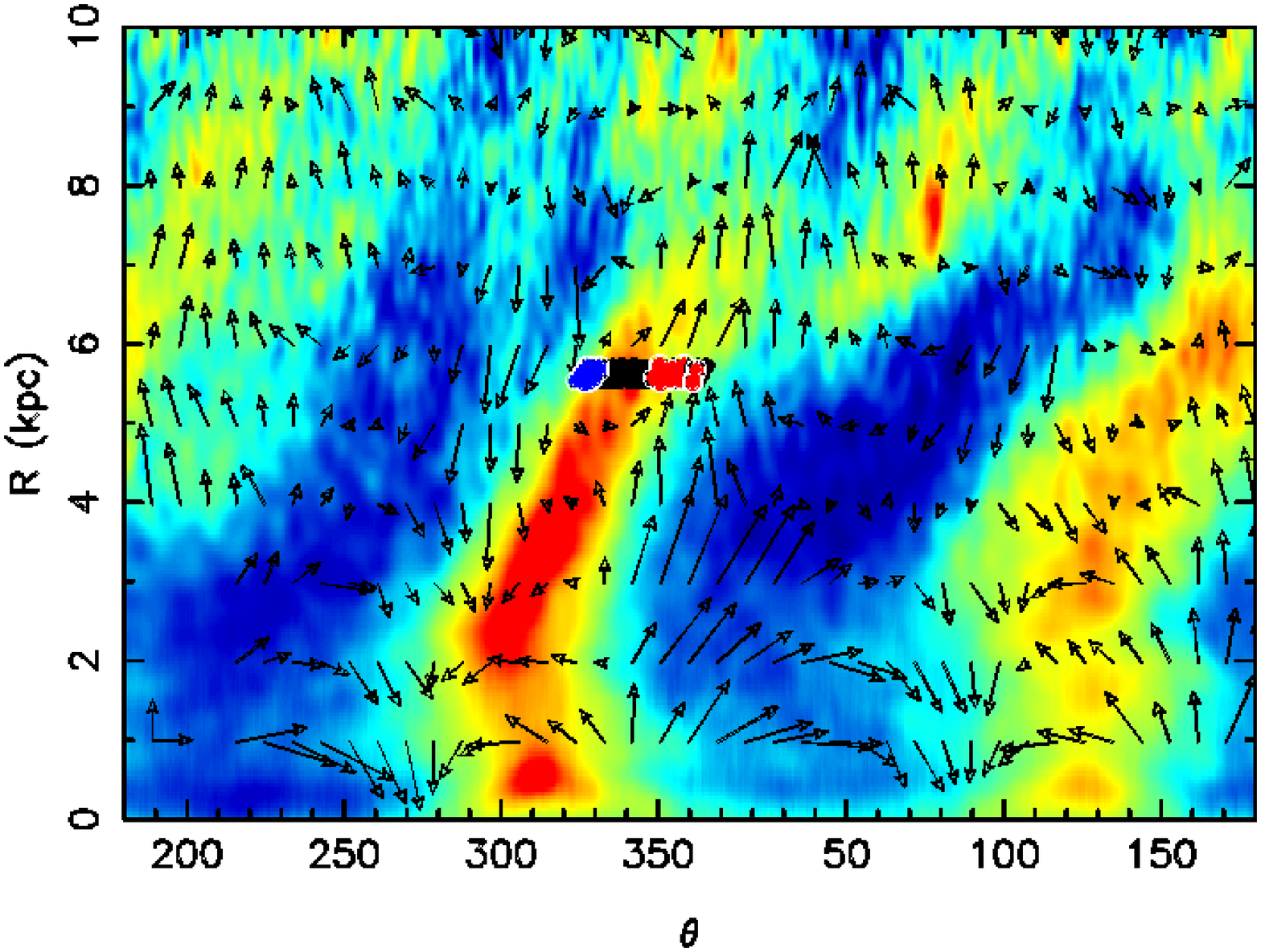} }
\resizebox{0.33\columnwidth}{!}{ \includegraphics{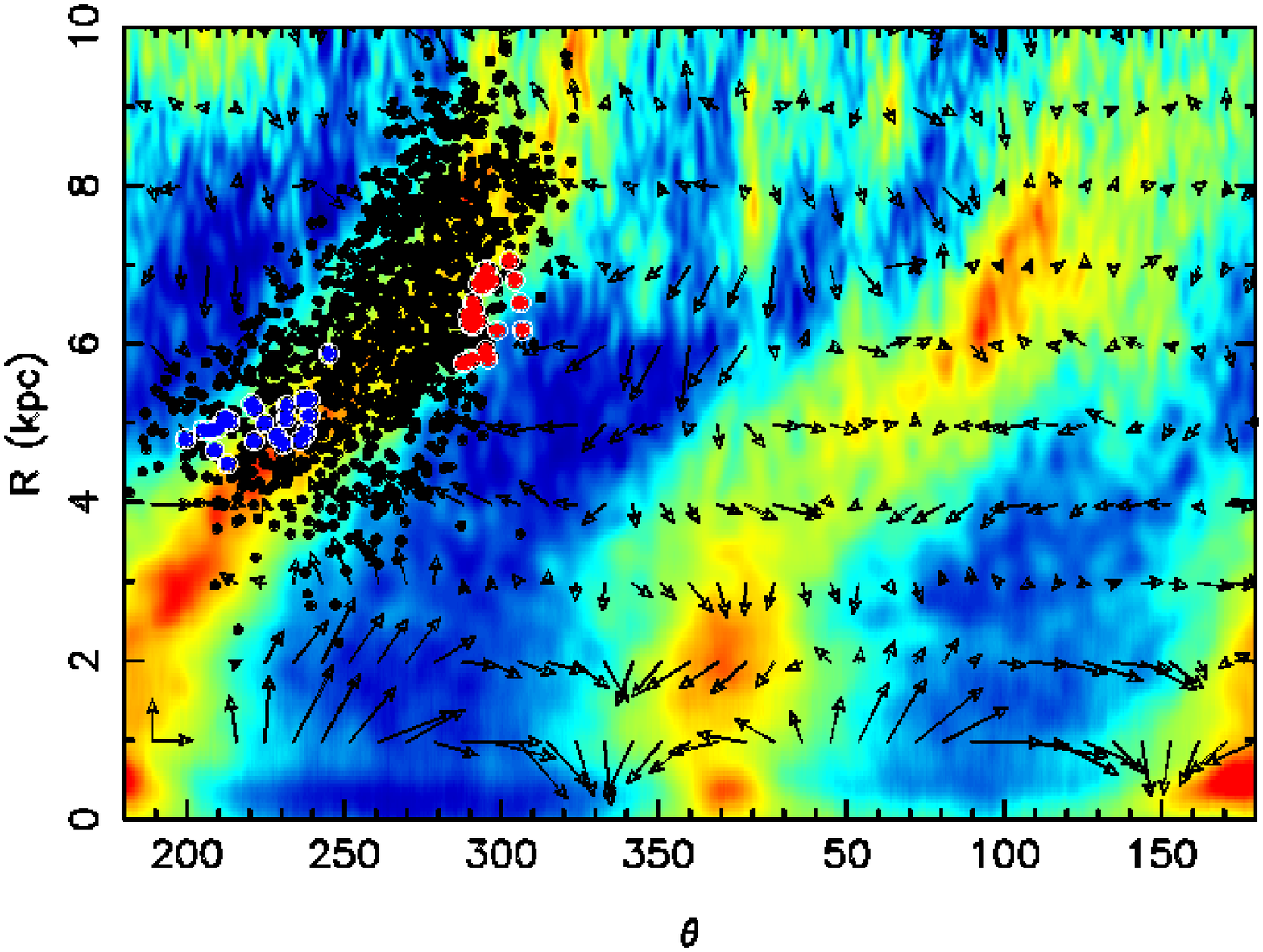} }
\caption{The normalised density map in the $\theta$ vs. R plane at $t=t_0-40$ Myr ({\it Left}), $t=t_0$ ({\it Middle})
and $t=t_0+40$ Myr ({\it Right}). Overplotted dots are selected particles around the spiral arm at $t=t_0$.
Arrows indicate the mean velocity field, The size of arrows in the bottom left corner corresponds to 20 km s$^{-1}$.
}
\label{fig:2}       % Give a unique label
\end{figure}

 As with the previous studies, we also find that the disk develops transient and recurrent spiral arms.
 We have computed a pattern speed of the spiral arm highlighted in Fig. \ref{fig:1},
and find a declining pattern speed with radius (right panel of Fig. \ref{fig:1}), so that pattern and
stars co-rotate at all radii, contrary to what the density wave theory predicts.
Since the arms wind up due to the differential rotation, the co-rotating spiral
arm is transient (see also \cite{wbs11}).
 We select a sample of star particles at 5.5-5.75 kpc within 20 deg away from the spiral 
arm and study their motion (Fig. \ref{fig:2}).
Figure highlights the evolution of the extreme positive (red dots) and negative (blue dots) migrators.
The 'positive' migrators are the particles that migrate towards the outer radii on the trailing side of the spiral
arm. They are trapped by the potential of the spiral arm, which accelerates them. The co-rotating nature of the spiral
arm ensures that during migration to outer radii, instead of passing through the spiral arm they remain on the trailing
side. Therefore, they continue to accelerate until the spiral arm is disrupted. The 'negative' migrators are particles that migrate towards the inner radii on the leading side of the spiral arm. They are decelerated as they become caught in the potential of the spiral arm, and because of the co-rotating nature they continue to decelerate on the leading side of
the spiral arm until the spiral arm is disrupted.
This means that they are accelerated or decelerated for longer, and results in
stronger migration along the arm. Since angular momentum is exchanged
via the co-rotating arm, there is little heating in their radial motion \cite{sb02}.

 We find similar transient co-rotating spiral arms in N-body simulations of non-barred disks and find
similar stellar motion around the spiral arms \cite{gkc11}.
We are studying further the spiral arms in our higher resolution simulations and endeavour to  find observational consequences of the co-rotating spiral arm.

% This illustrates the different motion that occurs on each side of the spiral arm.
% It is strikingly clear that negative migrators are captured by the spiral arm on
%the leading side of the arm, and the positive migrators dragged by the arm
%on the trailing side. Since the spiral arm is co-rotating, the stars stay in the
%vicinity of the arm, even after they migrate to the inner or outer region. T

%

%%% Fig. 1
%
%%% Fig. 2

\end{document}